# THERMAL RESILIENCE OF SUSPENDED THIN-FILM LITHIUM NIOBATE ACOUSTIC RESONATORS UP TO 550 °C

*Mihir Chaudhari, Naveed Ahmed, Vivek Tallavajhula, Joshua Campbell, Yinan Wang, Ziran Du, and Ruochen Lu*

Department of Electrical and Computer Engineering, The University of Texas at Austin, USA

## ABSTRACT

This paper reports a suspended thin-film lithium niobate (LN) piezoelectric resonator platform surviving high annealing temperatures of 550 °C, among the highest temperature at which the thermal resilience of suspended LN resonators is studied. Acoustic resonators are built on 600 nm thick transferred stoichiometric LN on silicon wafers with 40 nm thick platinum (Pt) electrodes, selected for high temperature operation. The fabricated resonators are first annealed at 250 °C, and the anneal temperature is incrementally increased to 550 °C after 7 rounds of annealing. The annealing is shown to upshift resonant frequencies and can increase the quality factor ($Q$), within a temperature range, before it gradually damages the device performance. This work presents promising results for using the suspended thin-film LN platform for resonators, sensors, and transducers in harsh thermal environments.

## KEYWORDS

Lithium niobate, acoustic resonator, anneal, piezoelectric device, high-temperature device.

## INTRODUCTION

Miniature piezoelectric acoustic devices have found use in various applications such as sensors, resonators, and transducers. Acoustic devices are an important foundation for sensors due to their high sensitivity to mechanical quantities such as force, pressure, and acceleration [1] [2]. Acoustic devices also have radio frequency applications due to their compact size and lower loss compared to electromagnetic solutions [3]. Ultrasound transducers based on miniature piezoelectric devices are commonly used for non-destructive testing in nuclear power plants, steel, plastic, and glass manufacturing processes, and in the aerospace field [4][5].

Amongst various applications, high-temperature acoustic devices have recently been drawing attention, especially in commonly used piezoelectric platforms such as lead zirconate titanate (PZT), aluminum nitride (AlN), scandium aluminum nitride (ScAlN), and lithium niobate (LN) [6] [7] [8] [9] [10]. A comparison of these piezoelectric platforms is given in Table 1. Compared to other platforms, LN is a strong piezoelectric platform for sensors, resonators, and transducers as it possesses higher figures of merit. Additionally, LN exhibits a high Curie temperature of 1200 °C. Thus, LN displays potential for use in harsh thermal environments.

However, the thermal resilience of the suspended thin-film LN acoustic platform, which supports bulk acoustic waves (BAWs) and Lamb wave resonators (LWRs), has not been studied up to the platform's temperature limit [11] [12] [13] [14]. The unique properties of suspended thin-film LN is especially important for high temperature operation, because different thermal expansion coefficients between the electrodes, the piezoelectric layer, and the supporting substrate are expected to break the rather fragile structure due to thermal mismatch [15], thus experimental demonstration on the performance of thin-film LN at high temperatures, and consequently methods to mitigate such issues could be useful for developing high-temperature efficient acoustic platforms.

| Material | Curie T (°C) | E-Field | Mode | Resonator FoM $e^2/(\varepsilon \cdot c)$ | Sensor/Transducer FoMs $e$ (C/m²) | $e/\varepsilon$ (GV/m) | $e/(\varepsilon \cdot \tan\delta)^{1/2}$ $(10^6(J/m^3)^{1/2})$ |
|---|---|---|---|---|---|---|---|
| PZT 5A | 350 | TFE | LE | 0.3% | 18.7 | 1.30 | 1.1 |
| AlN | 1150 | TFE | LE | 0.96% | 0.59 | 0.65 | 0.62 |
| AlN | 1150 | TFE | TE | 6.0% | 1.47 | 1.61 | 1.54 |
| Al₀.₇Sc₀.₃N | 1100 | TFE | LE | 0.84% | 0.70 | 3.75 | 2.70 |
| Al₀.₇Sc₀.₃N | 1100 | TFE | TE | 9.7% | 2.38 | 1.27 | 0.92 |
| X-cut LiNbO₃ | 1200 | LFE | LE | 31.8% | 4.61 | 13.24 | 6.85 |

*Table 1: Comparison between LN and other acoustic platforms (PZT, AlN, and ScAlN).*

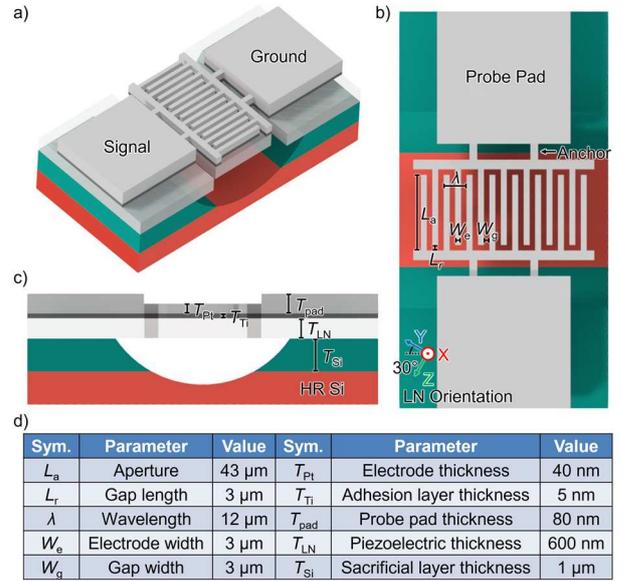

| Sym. | Parameter | Value | Sym. | Parameter | Value |
|---|---|---|---|---|---|
| $L_a$ | Aperture | 43 µm | $T_{Pt}$ | Electrode thickness | 40 nm |
| $L_r$ | Gap length | 3 µm | $T_{Ti}$ | Adhesion layer thickness | 5 nm |
| $\lambda$ | Wavelength | 12 µm | $T_{pad}$ | Probe pad thickness | 80 nm |
| $W_e$ | Electrode width | 3 µm | $T_{LN}$ | Piezoelectric thickness | 600 nm |
| $W_g$ | Gap width | 3 µm | $T_{Si}$ | Sacrificial layer thickness | 1 µm |

*Figure 1: (a) Mock-up view of suspended resonator for high-temperature test with (b) top view, (c) cross-sectional side view, and (d) table summarizing key dimensions.*

In this work, we study the effects on thin-film LN acoustic resonators as they are subjected to increasing temperatures. Studying the thermal resilience of the platform will enable new opportunities and applications for improved high operating temperature resonators, transducers, and sensors.

## DEVICE DESIGN AND FABRICATION

Thin-film LN LWR devices are designed and fabricated for high temperature study. Fundamental symmetric (S0) mode resonators are designed for 600 nm X-cut stoichiometric LN. Stoichiometric LN, which has a near 1:1 ratio of lithium and niobium, is reported to be better suited than congruent LN for high temperature applications, which has a slightly lower ratio of lithium to niobium [16] [17]. The devices are arranged with a 30° in-plane rotation for maximizing piezoelectric coupling [18]. The design and parameters are detailed in Fig. 1. Platinum (Pt) was chosen as the electrode metal because of its high melting point and stable properties compared to other electrode metals [19], and a thin layer of titanium (Ti) is included as an adhesion layer [20].

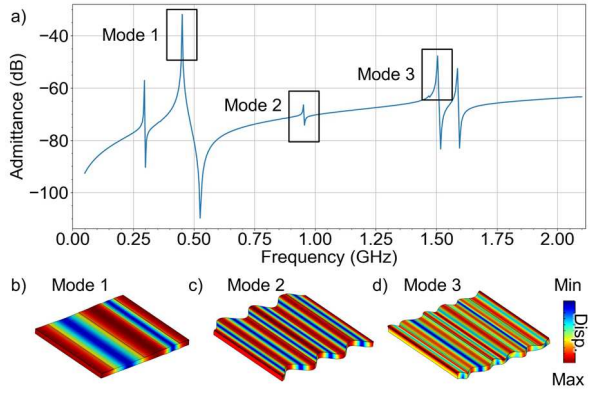

*Figure 2: Simulated device (a) admittance, and (b)-(d) mode shapes of the key modes of interest. Mode 1 (b) is the fundamental symmetric S0 mode, mode 2 (c) is the higher-order fundamental shear horizontal SH0 mode, and mode 3 (d) is a higher-order S0 mode.*

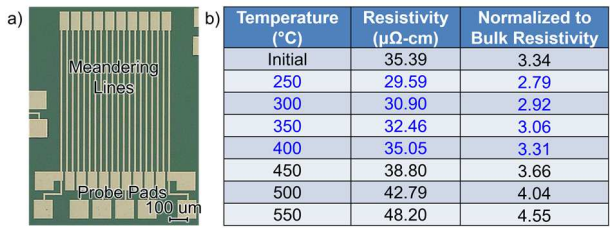

*Figure 3: (a) Meandering line structure for characterizing Pt resistivity (b) up to 550 °C, with rounds having a lower resistivity than the initial round highlighted in blue.*

| Temperature (°C) | Resistivity (µΩ-cm) | Normalized to Bulk Resistivity |
|---|---|---|
| Initial | 35.39 | 3.34 |
| 250 | 29.59 | 2.79 |
| 300 | 30.90 | 2.92 |
| 350 | 32.46 | 3.06 |
| 400 | 35.05 | 3.31 |
| 450 | 38.80 | 3.66 |
| 500 | 42.79 | 4.04 |
| 550 | 48.20 | 4.55 |

The resonator is simulated with COMSOL finite element analysis, and three main acoustic modes are selected for the study (Fig. 2). Mode 1 is the fundamental symmetric S0 mode, while modes 2 and 3 are higher-order fundamental shear horizontal (SH0) and higher-order S0 modes. To include a wider range of devices, additional resonators are designed with a parameter sweep.

The resonators are fabricated in a cleanroom with standard surface micromachining techniques. First, release etch windows are defined with photolithography and etched with ion milling. Next, the electrodes are defined with photolithography, and 5 nm of Ti and 40 nm of Pt are deposited onto the LN with electron beam metal evaporation. Then, the unwanted metal is removed via a lift-off process. Finally, the resonators are suspended with a xenon difluoride (XeF$_2$) etch, which etches the sacrificial silicon layer.

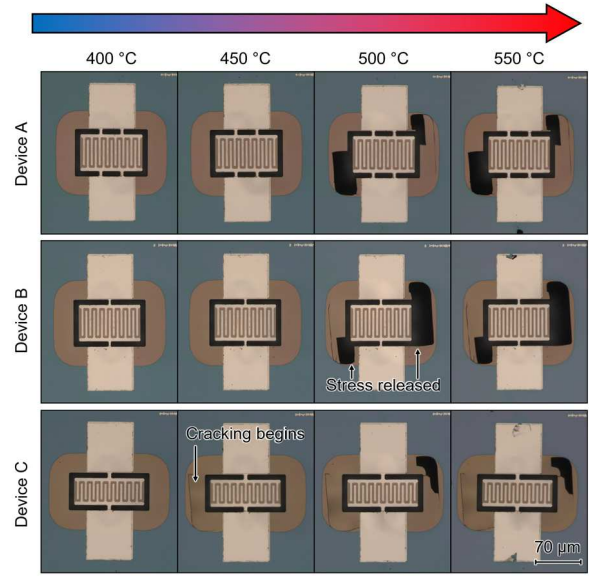

*Figure 4: Device microscopic images after thermal cycles of 400 to 550 °C.*

In addition to the resonators, meandering line resistivity structures (Fig. 3(a)) are fabricated on the same substrate to monitor the metal resistivity after annealing rounds. The resistivity structures are composed of 40 nm of Pt on top of 5 nm of Ti, matching the resonator electrodes. The meandering line structure enables testing resistivity in a compact form factor, ensuring the resistance is large enough to be measured by DC probes. Multiple probe pads enable measuring the resistance of multiple paths, which will allow average resistance measurements.

## ANNEALING RESULTS

After device fabrication, a cycle of device characterization and annealing is performed from 250 °C to 550 °C, which is the temperature limit of the annealing oven, in increments of 50 °C. The vacuum anneal begins by flooding the chamber with nitrogen, then setting the chamber to high vacuum (~µTorr). A ramp rate of 100 °C/hr was used to heat to/cool from the target temperature, where the target temperature was held constant for 10 hours. After each annealing round, the sample goes through an inspection and characterization procedure. First, optical microscope images of the devices post-anneal are taken. Next, the metal resistivity is characterized by measuring the resistance of the meandering lines. Finally, the piezoelectric resonator response is characterized by a Keysight vector network analyzer (VNA) measurement.

The optical microscope images give a qualitative description of the annealing effects. From Figure 4, it can be seen that at 500 °C, the LN film surrounding the devices begins to crack as indicated by the black spots around the resonators. The cracking is caused by stress release in the LN. Because of differences in the coefficient of thermal expansion between LN, the supporting silicon substrate, and the electrodes, the LN experiences compressive stress during annealing and tensile stress when cooling to room temperature [21]. Although the cracks appear in the regions where the resonators are suspended, the resonators do not collapse. At 550 °C the metal electrodes become more brittle, and are more easily scraped by electrical probes.

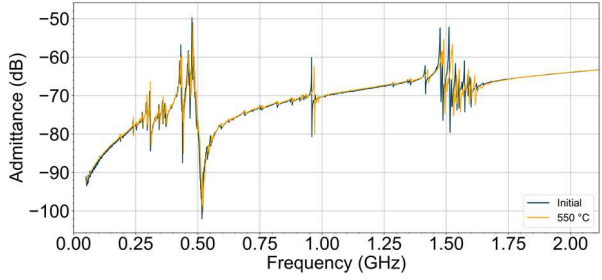

*Figure 5: Wideband admittance before annealing and after a 550 °C anneal. A frequency upshift is observed.*

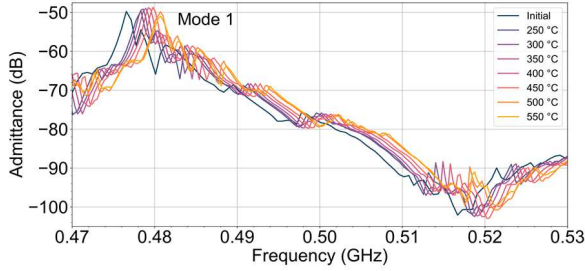

| Temp. (°C) | Δ Resistivity (µΩ-cm) | f (MHz) | Extracted Q | Δ Q |
|---|---|---|---|---|
| Initial | 0 | 467.777 | 458.0 | 0 |
| 250 | -5.8 | 478.278 | 430.7 | -27.3 |
| 300 | -4.49 | 478.540 | 435.2 | -22.8 |
| 350 | -2.39 | 478.911 | 449.7 | -8.3 |
| 400 | -0.34 | 479.366 | 452.6 | -5.4 |
| 450 | 3.41 | 479.478 | 453.5 | -4.5 |
| 500 | 7.40 | 480.668 | 348.2 | -109.8 |
| 550 | 12.81 | 480.724 | 329.6 | -128.4 |

*Figure 6: Mode 1 (S0) admittance and resonance parameters. The frequency increases with every annealing round. While the Q remains below the initial value, it begins to increase after an initial decrease while the metal resistivity is below its initial value.*

The metal resistivity derived from the meandering lines is a direct indicator of resistor electrode quality. The resistivities are shown in Fig. 3(b), with a differential fitting using measurement of resistance in different pads. A higher metal resistivity will lead to a higher series resistance and degrade the performance of the resonator. The initial anneal at 250 °C reduces the metal resistivity as the metal lattice defects are cleansed [20]. In subsequent annealing rounds, the metal resistivity begins increasing, and exceeds the initial resistivity after the 450 °C anneal. While an initial anneal removes defects, subsequent anneals can introduce more defects such as holes and grain coarsening [20].

The electrical response is characterized by admittances measured with a VNA. Parameters of resonance features are extracted from curve fitting. Figure 5 shows the wideband response of the resonator (Device A). Spurious modes account for the differences between the measured admittance and the simulated admittance. The admittances of the three modes after each annealing round are shown individually in Figures 6, 7, and 8 with tables for extracted resonance parameters. Additionally, the changes in resistivity and changes in Q are tabulated to compare with the initial pre-anneal conditions.

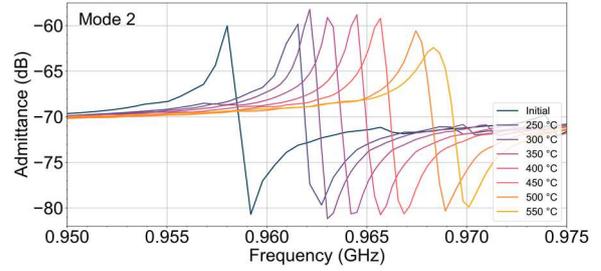

| Temp. (°C) | Δ Resistivity (µΩ-cm) | f (MHz) | Extracted Q | Δ Q |
|---|---|---|---|---|
| Initial | 0 | 958.149 | 1398.4 | 0 |
| 250 | -5.8 | 961.398 | 1528.7 | 130.3 |
| 300 | -4.49 | 962.154 | 2008.4 | 610.0 |
| 350 | -2.39 | 963.219 | 1884.8 | 486.4 |
| 400 | -0.34 | 964.473 | 1850.3 | 451.9 |
| 450 | 3.41 | 965.679 | 1678.5 | 280.1 |
| 500 | 7.40 | 967.691 | 1232.7 | -165.7 |
| 550 | 12.81 | 968.651 | 851.2 | -547.2 |

*Figure 7: Mode 2 (higher-order SH0) admittance and resonance parameters. Annealing rounds where the Q is above its initial value are highlighted in blue. The Q value exceeds the initial value when the metal resistivity is below its original value, and at 450 °C when the metal resistivity is above its original value.*

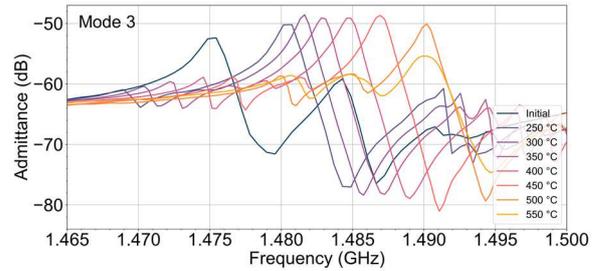

| Temp. (°C) | Δ Resistivity (µΩ-cm) | f (GHz) | Extracted Q | Δ Q |
|---|---|---|---|---|
| Initial | 0 | 1.475 | 717.0 | 0 |
| 250 | -5.8 | 1.481 | 969.8 | 252.8 |
| 300 | -4.49 | 1.482 | 1162.6 | 445.6 |
| 350 | -2.39 | 1.483 | 1096.4 | 379.4 |
| 400 | -0.34 | 1.485 | 1072.4 | 355.4 |
| 450 | 3.41 | 1.487 | 1115.8 | 398.8 |
| 500 | 7.40 | 1.490 | 1058.9 | 341.9 |
| 550 | 12.81 | 1.490 | 631.1 | -85.9 |

*Figure 8: Mode 3 (higher-order S0) admittance and resonance parameters. Annealing rounds where the Q is above its initial value are highlighted in blue. The Q exceeds the initial value while the metal resistivity is below its original value, and two rounds afterwards when the resistivity is above its original value. Frequency upshifting stops after 500 °C and 1.490 GHz.*

The three modes follow general trends while having unique differences after annealing. Every mode experiences a frequency upshift after an annealing round, except for mode 3 which reaches a frequency limit at 500 °C. The resonant frequency increases because annealing builds stress in the LN, which increases the speed of sound of the acoustic wave. Since the acoustic wavelength remains the same, as determined by electrode dimensions, the resonant frequency will increase. Despite a lower metal resistivity, mode 1 does not experience an increase in Q compared to its initial value. Modes 2 and 3 have Qs that exceed the initial value for annealing rounds where the metal resistivity is lower than the original. Additionally, mode 2 has a *Q* value above the original after an anneal at 450 °C, and mode 3 has *Q* values above the original after anneals at 450 and 500 °C, where the metal resistivity is larger than its original value. At 550 °C, all modes begin to degrade as the *Q* decreases. Despite degradations in performance with successive annealing rounds up to 550 °C, the resonators still have an electrical response, demonstrating the thermal resilience of the suspended thin-film LN platform. Additionally, annealing can be introduced as a post-fabrication step to enhance the *Q* and increase the frequency of a resonator.

## CONCLUSION

The thin-film suspended LN platform can survive annealing temperatures up to 550 °C. Due to metal quality and *Q* degradation, the ideal annealing temperature for improving the device *Q* will be below 550 °C, although not all modes will experience an improvement. All modes have a general trend of increasing their resonant frequency. However, the resonators survive and are still operational at 550 °C, proving the strong thermal resilience of the thin-film suspended LN platform. Further study with higher annealing temperatures will determine the ultimate thermal resilience limit of suspended thin-film LN platforms.

## ACKNOWLEDGEMENTS

The authors would like to thank DARPA HOTS for funding support and Dr. Todd Bauer for helpful discussions.

## CONTACT

*Mihir Chaudhari, mihirc@utexas.edu